\documentclass[10pt]{article}  
\usepackage{fullpage}

\usepackage{mathptmx} 
\usepackage{amsmath} 
\usepackage{graphicx}

\title{\LARGE \bf Classical General Relativity Effects to Second Order in Mass, Spin, and Quadrupole Moment}

\author{J.~R.~Arce-Gamboa and 
F.~Frutos-Alfaro\thanks{School of Physics, Universidad de Costa Rica}}  

\begin{document}

\maketitle


\begin{abstract}

In this contribution, we calculate the light deflection, perihelion shift, 
time delay and gravitational redshift using an approximate metric that contains 
the Kerr metric and an approximation of the Erez-Rosen spacetime.   
The results were obtained directly using Mathematica \cite{Wolfram}. 
The results agree with the ones presented in the literature, but they are 
extended until second order terms of mass, angular momentum and mass 
quadrupole. The inclusion of the mass quadrupole is done by means of the 
metric; no expansion of the gravitational potential as in the parameterized 
post-Newtonian is required. 

\end{abstract}


\section{Introduction}
\label{intro}

Einstein's General Theory of Relativity is a metric theory of gravity that
relates the mass-energy content of the universe with the space-time curvature 
through the Einstein field equations:

\begin{equation}
\label{EE}
R_{ab} - \frac{1}{2} g_{ab} R + g_{ab} \Lambda = 8 \pi \frac{G}{c^4} T_{ab} ,
\end{equation}

\noindent
where the right-hand side of this equation depends on the stress-energy tensor 
$ T_{ab} $ which describes the mass-energy sources of gravitational fields, 
and the left-hand side depends on the metric elements $ g_{ab} $ which describe 
the space-time curvature. $ R_{ab} $ are the Ricci tensor components, $ R $ is 
the scalar curvature, and $ \Lambda $ is the cosmological constant. 
In this article, the geometrical units are employed, so that $ G = c = 1 $.
The cosmological constant is set $ \Lambda = 0 $.
  
In 1916, Karl~Schwarzschild discovered a solution to the Einstein field 
equations in vacuum, suitable for describing the spacetime in the empty space 
surrounding a spherical, static object \cite{Schwarzschild}. Ever since then, 
this metric has been used to describe a wide range of phenomena, including 
light deflection close to a massive star, planetary precession of the 
perihelion, time delay and gravitational redshifts for weak fields. 
G.~Erez and N.~Rosen introduced the effects of mass quadrupole $q$ as exact 
solution in 1959 \cite{Carmeli,ER}. This derivation had some errors which were 
corrected by Doroshkevich et al. \cite{DZ}, Winicour et al. \cite{WJN} and 
Young and Coulter \cite{YC}.  
The exact solution for a rotating black hole (BH) could only be solved as late 
as 1963 by Roy~P.~Kerr \cite{Kerr}. There are exact solutions containing 
the Erez-Rosen and Kerr features, such spacetimes are cumbersome. 
A new approximate metric representing the spacetime of a rotating deformed 
body is obtained by perturbing the Kerr metric to include up to the second 
order of the quadrupole moment \cite{FA2}. This kind of approximations is valid 
because the quadrupole moment is small generally for a variety of 
astrophysical objects. Observing the Spin of rotating BH is possible by 
measuring the orbital angular momentum of light propagating around it, 
as well as BH shadow circularity analysis \cite{Tamburini}.

In the literature, calculations that include the mass quadrupole are only done
using (parameterized) post Newtonian metrics. To introduce the mass quadrupole,
the gravitational potential is expressed as a multipolar expansion
\cite{MN,Paez, Quevedo, Richter0,Richter1,Richter2,Richter3}. 
In our calculation we perform no such expansion of the gravitational potential.
The quadrupole parameter is introduced from the metric.

Now, it is possible to do such calculation in a straightforward manner using 
software like Mathematica. In this contribution, we present the 
results of light deflection, perihelion shift, time delay and gravitational 
redshift using this software. The results were compared with the ones obtained 
from the Reduce software.        

This paper is organized as follows. The classical tests of general relativity
are described in section \ref{classicalTests}. The parameterized post-Newtonian
formalism is introduced in section \ref{PPN}. The approximate metric with three 
parameters ($ M, \, J = m a, \, q $) is described in section \ref{Metric}. 
The metric potentials are expanded in a Taylor series up to second order of 
$ J $, $ M $ and $ q $. The resulting metric is transformed into a 
Hartle-Thorne form. In section \ref{deflection} we calculate the angle of the 
deflection of light in traveling in the equatorial plane of our metric. 
In section \ref{perihelion}, we present the necessary calculations to obtain 
the angle of Precession of the perihelion of the orbit of a planet in 
the presence of a space-time described by our metric. In section \ref{delay}, 
we calculate the time delay of light traveling between two points and in 
section \ref{redshift}, the expression for the gravitational redshift 
in two different positions in our space-time is obtained. 
The Mathematica notebook is available upon request. 
Our concluding remarks are presented in the last section.  


\section{The Classical Tests}
\label{classicalTests}

In the solar system, most of the Newtonian mechanics predictions are in good 
agreement with observations. However, there are a few situations where general 
relativity (GR) is positioned as a more precise theory.  Traditionally, 
they are Mercury's perihelion precession, the light deflection by the Sun, 
the gravitational redshift of light and the time delay of light. 

Mercury's perihelion precession is the first classical test and was first 
noted by Le Verrier in 1859. In this phenomenon, classical contributions such 
as the  planetary perturbations influence \cite{Lo, Ludl}, yet it remains 
a discrepancy of $ 42.7 '' $ per century. The  contributions  from GR reports 
a value of $ 42.95 '' $ per century. During the 1960's and 1970's there was 
a considerable controversy on the importance of the contribution of the solar 
oblateness mass quadrupole $ J_2 $ on the perihelion precession. 
This discussion has relaxed as the value of the solar quadrupole has been 
inferred to be small, on the order of $ J_2 = (2.25 \pm 0.09) \times 10^{-7} $ 
\cite{will2018theory, Ludl}.  Using this value, it has been estimated that 
the contribution to the precession from the solar oblateness is of 
$ 0.0286 \pm 0.0011 '' $ per century. Yet, its importance can not be specified 
until a reliable value of the quadrupole is known. The second test, the light 
deflection due to the massive body of the Sun, was famously first observed 
during the Eddington's expedition in 1919 with a high degree of inaccuracy, 
but it was not observed with precision until the 70's using radio wave 
interferometry.  By this time, it was reported that the mean gravitational 
deflection was $ 1.007 \pm 0.009 $ times the value predicted by GR \cite{Ludl}. 
The deflection caused by the solar oblateness can be treated as a small 
correction. Typically, it could modify the path of ray of light in 
$ 0.2 \, \mu $ arcseconds. Other physical property that influences light 
deflection is the Sun's angular momentum, as it has been calculated that 
the Sun's amount of $ L \approx 2 \times 10^{41} \, {\rm kg \, m^2/s} $ can be 
responsible for a deflection of $ 0.7 \, \mu $ arcseconds \cite{Epstein}.

The third test, the gravitational redshift, measures the wavelenght shift 
between two identical clocks placed at rest at different positions in 
a gravitational field.  This was the first test to be proposed by Einstein, 
and was first tested by Pound, Rebka and Snider in the 1960s, as they measured 
the gamma radiation emitted by $ {}^{57}$Fe , as they ascended or descended 
the Jefferson Physical Laboratory tower \cite{Ludl}. The fourth test, 
the gravitational time delay, was classified as such by Will and was first 
observed by Shapiro in 1964 when he discovered that a ray of light propagating 
in the gravitational field of a massive body will take more time traveling 
a given distance, than if the field were absent \cite{Ludl}. Gravitational 
time delay can be observed by measuring the round trip of a radio signal 
emitted from Earth and reflected from another body, such as another planet or 
a satellite.  To properly measure the effect, it is necessary to do 
a differential measurement in the variations in the round trip as the target 
object moves through the sun's gravitational field.  This task is particularly 
difficult as it involves taking into account the variations in the round trip 
as a result of the orbital motion of the target relative to Earth 
\cite{will2018theory}.

Another ideal probe for testing GR is the massive black hole (MBH) 
located in a bright and very compact astronomical radio source called
Sgr A* at the center of the Milky Way at a distance $ R_0 \approx 8 $ kpc
and with a mass $ M_{\bullet} \approx 4 \times 10^6 M_{\odot} $.  This MBH is 
surrounded by the highly elliptical star S2 whose motion has been an important 
subject of study in the literature \cite{Zucker, Gravity}.  It has been 
determined that S2 has a semi-major axis $ a = 8122 \pm 31 $ mas and 
an eccentricity $ e = 0,88466 \pm 0,000018 $, and so is possible to make 
an estimate of the contributions of the mass of the MBH to the orbit 
precession and the gravitational redshift and compare them with the values 
reported in the literature.


\section{The Parametrized Post-Newtonian Formalism}
\label{PPN}

Although it has been very successful when compared with direct observations, 
GR is just one of many metric theories of gravity, and all that distinguishes 
one metric theory from another is the particular way in which matter generates 
the metric. It is simple to perform a comparison between metric theories in 
the slow-motion and weak-field limit, since all of their results must agree 
with Newtonian phyisics.

The parametrized post-Newtonian (PPN) formalism is a device that allows 
the comparison between different theories of gravitation and experiments.  
It is motivated by the advent of alternative theories of gravitation other 
than GR during the second half of the twentieth century. 
It has provided a common framework to quantify deviations from GR which are 
small in the post-Newtonian order.
 
As the various theories of gravitation involve mathematical objects such as 
coordinates, mass variables and metric tensors, PPN formalism is provided with 
a set of ten parameters which describe the physical effects of these theories.  
The so called Eddington-Robertson-Schiff parameters $ \gamma $ and $ \beta $ 
are the only non-zero parameters in GR, hence they are significant in the study 
of classical tests. $ \beta $ measures whether gravitational fields do 
interact with each other, while $ \gamma $ quantifies the space-curvature 
produced by unit rest mass, and both their values is one in GR \cite{Ludl}.

In this context, it is very important to mention Gaia, the ESA space 
astrometry mission launched in late 2013. Through its detectors, it will 
perform Eddington-like experiments through the comparison between the pattern 
of the starfield observed with and without Jupiter. For this purpose, it is 
vital to have a formula relevant for the monopole and quadrupole light 
deflection for an oblate planet. These results will provide a new independent 
determination of $ \gamma $ and evidence of the bending effect of the mass 
quadrupole of a planet \cite{Crosta2006, Crosta2007}. It is currently accepted 
that $ | 1 - \gamma | $ is less than $ 2 \times 10^{-5} $.

It is also relevant to highlight the use of radiometric range measurements to 
the MESSENGER spacecraft in orbit around Mercury to estimate the precession of 
Mercury's perihelion. Knowing a suitable relationship between this classical 
test and the quadrupole allows to decouple $ \beta $ and the solar quadrupole 
$ J_2 $ to yield  $ (\beta - 1) = (- 2.7 \pm 3.9) \times 10^{-5} $ \cite{Park}. 
It has been conjectured that there is another additional contribution to 
the perihelion advance from the relativistic cross terms in the post-Newtonian 
equations of motion between Mercury's interaction with the Sun and with 
the other planets, as well from the interaction between Mercury's motion and 
the gravitomagnetic field of the moving planets.  These effects are planned to 
be detected by the BepiColombo mission, launched in late 2018 \cite{Will}. 

There have been several papers that have quantified the contributions to the 
classical tests from various objects in the solar system. Detection and precise 
measurement of the quadrupolar deflection of light by objects in the solar 
system, at the level of a microarcsecond positional accuracy,  is important as 
it will allow the experimental observation of a wide range of physical 
phenomena that will allow to test GR in a velocity and 
acceleration-independent-regime.  There are research lines that study 
the effects related to the motion of planets such as the appearance of 
a gravitational field due to the mass dipole and methods to properly measure 
the quadrupole of the planets that compensate for the effects due to their 
movements \cite{Kopeikin}. The values shown in Table \ref{table:1} illustrate 
the maximal magnitudes of the various gravitational effects due to the Sun and 
the planets at which the gravitational light deflection from that body should 
still be accounted for to attain a final accuracy of 1 $\mu$as. Here, 
Second Order: PN is the post-Newtonian effect due to the spherically symmetric 
field of each body, Rotation accounts for the field caused by the rotational 
motion of the bodies, Fourth Order: PPN is the post-post Newtonian effect due 
to the mass, and  Quadrupole: PN is the effect caused by the mass quadrupole 
\cite{klioner}.

\begin{table}[ht!]
\centering
\begin{tabular}{|c c c c c|} 
\hline
Stellar Object & Second Order: PN ($\mu$as) &Rotation ($\mu$as) & 
Fourth Order: PPN ($\mu$as) & Quadrupole: PN ($\mu$as)  \\ [0.5ex] 
\hline\hline
Sun & $1.75 \times 10^{+6}$ &0.7 & 11 & $\sim$ 1 \\ 
Mercury & $83$ & - & - & -\\ 
Venus & $493$ & - & - & -\\ 
Earth & $574$ & - & - & $0.6$\\ 
Mars & $116$  & -& - & $0.2$\\ 
Jupiter & $16270$  & 0.2& - & $240$ \\ 
Saturn & $5780$ & - & -& $95$\\ 
Uranus & $2080$ & -& - & $8$\\ 
Neptune & $2533$ & -& - & $10$\\ [1ex] 
\hline
\end{tabular}
\caption{Order of magnitude of the contributions PN, PPN, $ {\rm PN}_{\rm Q} $ 
and $ {\rm PN}_{\rm R} $ to the deviation angle of a light ray grazing the solar 
limb as predicted by GR \cite{klioner}.}
\label{table:1}
\end{table}

Table \ref{table:2} shows the values of the contributions to the gravitation 
delay of a radio signal as it is measured from the Earth \cite{Paez}.  

In this formalism, the gravitational potential of an axially symmetric body can be written in the following form \cite{FA3}

\begin{eqnarray}
\label{uppn}
\frac{\cal U}{c^2} = \frac{G M}{c^2 r} + \frac{G q}{c^2 r^3} P_2(\cos{\theta}) .
\end{eqnarray}

In this paper, we will consider up to second order in the PPN formalism. 

\begin{table}[ht!]
\centering
\begin{tabular}{|c c c c c|} 
\hline
Stellar Object & Second Order: PN (ns)  & Rotation (ns) & 
Fourth Order: PPN (ns) & Quadrupole: PN (ns) \\ [0.5ex] 
 \hline\hline
Sun & $1.1946 \times 10^{+5}$ &$7.894 \times 10^{-3}$ & $1.8091 \times 10^{+1}$ 
& $5.4179 \times 10^{-2}$ \\ 
Mercury & $3.6722 \times 10^{-2}$ &$1.2965 \times 10^{-11}$ 
& $2.4716 \times 10^{-8}$ & - \\ 
Venus & $4.5932 \times 10^{-1}$ &$9.8968 \times 10^{-11}$ 
& $3.9434 \times 10^{-7}$ & - \\ 
Mars & $6.8286 \times 10^{-2}$ &$1.9160 \times 10^{-9}$ 
& $4.1215 \times 10^{-8}$ & $6.2437 \times 10^{-6}$ \\ 
Jupiter & $1.8402 \times 10^{+2}$ &$1.9543 \times 10^{-4}$ 
& $6.6439 \times 10^{-3}$ & $1.3870 \times 10^{-1}$ \\ 
Saturn & $6.0039 \times 10^{+1}$ &$4.1924 \times 10^{-5}$ 
& $1.6942 \times 10^{-3}$ & $4.6307 \times 10^{-2}$ \\ 
Uranus & $1.0594 \times 10^{+1}$ &$1.3220 \times 10^{-6}$ 
& $5.0213 \times 10^{-4}$ & $5.1645 \times 10^{-3}$ \\ 
Neptune & $1.2993 \times 10^{+1}$ &$3.3923\times 10^{-6}$ 
& $1.0775 \times 10^{-3}$ & $2.0365 \times 10^{-3}$ \\ [1ex] 
\hline
\end{tabular}
\caption{Order of magnitude of the contributions PN, PPN, and 
$ {\rm PN}_{\rm Q} $ to the gravitation delay of a radio signal grazing 
the solar limb and the planets predicted by GR using a PPN metric \cite{Paez}.}
\label{table:2}
\end{table}


\section{The Metric} 
\label{Metric}

The metric, we will employ to do the calculations was generated in a 
perturbative form using the Kerr spacetime as seed metric. This approximate 
rotating spacetime with quadrupole moment written in standard form is as 
follows \cite{FA1,FA2}:

\begin{eqnarray} 
\label{superkerr}
d{s}^2 & = & - \frac{\Delta}{{\rho}^2} 
[{\rm e}^{- \psi} dt - a {\rm e}^{\psi} \sin^2{\tilde{\theta}} d \phi]^2 
\nonumber \\
& + & \frac{\sin^2{\tilde{\theta}}}{{\rho}^2} 
[({\tilde{r}}^2 + a^2) {\rm e}^{\psi} d \phi - a {\rm e}^{- \psi} d t ]^2 
\nonumber \\
& + & {\rho}^2 {\rm e}^{2 \chi} \left(\frac{d {\tilde{r}}^2}{\Delta} 
+ d {\tilde{\theta}}^2 \right) ,
\end{eqnarray}

\noindent
where 

\begin{eqnarray}
\label{functions}
\Delta & = & {\tilde{r}}^2 - 2 M {\tilde{r}} + a^2 , \nonumber \\  
{\rho}^2 & = & {\tilde{r}}^2 + a^2 \cos^2{\tilde{\theta}} , \\
\psi & = & \frac{q}{{\tilde{r}}^3} P_2 + 3 \frac{M q}{{\tilde{r}}^4} P_2 , 
\nonumber \\ 
\chi & = & \frac{q}{{\tilde{r}}^3} P_2
+ \frac{1}{3} \frac{M q}{{\tilde{r}}^4} (5 P_2^2 + 5 P_2 - 1) \nonumber \\
& + & \frac{1}{9} \frac{q^2}{{\tilde{r}}^6} (25 P_2^3 - 21 P_2^2 - 6 P_2 + 2) , 
\nonumber \\
P_2 & = & \frac{1}{2} ({3\cos^2{\tilde{\theta}} - 1}) . \nonumber
\end{eqnarray}

This spacetime has three parameters, namely mass $ M $, spin, $ J = M a $ 
($ a $ as the Kerr rotation parameter) and $ q $, the mass quadrupole. 
It contains the Kerr and the Schwarzschild  metrics. This metric is 
an approximation to the Erez-Rosen metric ($ q^3 \sim 0 $).    

According to \cite{FA2}, the Taylor series up to second order of 
$ a, \, J, \, M $ and $ q $ gives 

\begin{eqnarray}
\label{postnewton}
g_{t t} & = & - \left(1 - 2 \frac{M}{\tilde{r}} 
+ 2 \frac{M a^2}{\tilde{r}^3} \cos^2{\tilde{\theta}} 
- 2 \frac{q}{\tilde{r}^3} P_2 - 2 \frac{M q}{\tilde{r}^4} P_2 
+ 2 \frac{q^2}{\tilde{r}^6} P_2^2 \right) \nonumber \\
g_{t \phi} & = & - 2 \frac{J}{\tilde{r}} \sin^2{\tilde{\theta}} \\
g_{\tilde{r} \tilde{r}} & = & 1 + 2 \frac{M}{\tilde{r}} + 4 \frac{M^2}{\tilde{r}^2} 
- \frac{a^2}{\tilde{r}^2} \sin^2{\tilde{\theta}} 
- 2 \frac{M a^2}{\tilde{r}^3} (1 + \sin^2{\tilde{\theta}}) 
- 4 \frac{M^2 a^2}{\tilde{r}^4} (2 + \sin^2{\tilde{\theta}}) \nonumber \\
& + & 2 \frac{q}{\tilde{r}^3} P_2 
+ \frac{2}{3} \frac{M q}{\tilde{r}^4} (5 P_2^2 + 11 P_2 - 1) 
+ \frac{2}{9} \frac{q^2}{\tilde{r}^6} (25 P^3_2 - 12 P^2_2 - 6 P_2 + 2) 
\nonumber \\
g_{\tilde{\theta} \tilde{\theta}} & = & 
r^2 \left(1 + \frac{a^2}{\tilde{r}^2} \cos^2{\tilde{\theta}} 
+ 2 \frac{q}{\tilde{r}^3} P_2 
+ \frac{2}{3} \frac{M q}{\tilde{r}^4} (5 P_2^2 + 5 P_2 - 1) 
+ \frac{2}{9} \frac{q^2}{\tilde{r}^6} (25 P_2^3 - 12 P_2^2 - 6 P_2 + 2) \right)
\nonumber \\
g_{\phi \phi} & = & {\tilde{r}}^2 \sin^2{\tilde{\theta}} 
\left(1 + \frac{a^2}{\tilde{r}^2} 
+ 2 \frac{M a^2}{\tilde{r}^3} \sin^2{\tilde{\theta}}
+ 2 \frac{q}{\tilde{r}^3} P_2 + 6 \frac{M q}{\tilde{r}^4} P_2 
+ 2 \frac{q^2}{\tilde{r}^6} P_2^2 \right) .
\nonumber
\end{eqnarray}

Now, in \cite{FA2} a transformation was found that converts this expanded 
metric (\ref{postnewton}) into the expanded Hartle-Thorne (HT) metric changing 
$ q \rightarrow M a^2 - q $ that included the second order in $ q $, it is

\begin{eqnarray}
\label{trans}
{\tilde{r}} & = & r \left[1 + \frac{M q}{r^4} f_1 + \frac{q^2}{r^6} f_2 
+ \frac{a^2}{r^2} \left({h_1} + \frac{M}{r} h_2 + \frac{M^2}{r^2} h_3 \right) 
\right] \\
{\tilde{\theta}} & = & \theta + \frac{M q}{r^4} g_1 + \frac{q^2}{r^6} g_2 
+ \frac{a^2}{r^2} \left({h_4} + \frac{M}{r} h_5 \right) , \nonumber
\end{eqnarray}

\noindent
where

\begin{eqnarray}
\label{functs}
f_1 & = & - \frac{1}{9} (1 + 4 P_2 - 5 P_2^2) \nonumber \\
f_2 & = & - \frac{1}{72} (43 + 24 P_2^2 - 40 P_2^3) \nonumber \\
g_1 & = & \frac{1}{6} (2 - 5 P_2) \cos{\theta} \sin{\theta} 
\nonumber \\
g_2 & = & \frac{1}{6} (2 - 5 P_2) P_2 \cos{\theta} \sin{\theta} \\
h_1 & = & - \frac{1}{2} \sin^2{\theta} \nonumber \\
h_2 & = & - \frac{1}{2} \sin^2{\theta} \nonumber \\
h_3 & = & 1 - 3 \cos^2{\theta} \nonumber \\
h_4 & = & - \frac{1}{2} \cos{\theta} \sin{\theta} \nonumber \\
h_5 & = & - \cos{\theta} \sin{\theta} . \nonumber 
\end{eqnarray}

the transformed metric components take the following form \cite{FA2} 

\begin{eqnarray} 
\label{components}
g_{tt} & = & - \left(1 - 2 U + 2 \frac{Q}{r^3} P_2 
- \frac{2}{3} \frac{J^2}{r^4} {(2 P_2 + 1)} 
+ 2 \frac{M Q}{r^4} P_2 + 2 \frac{Q^2}{r^6} P_2^2 \right) \nonumber \\
g_{t \phi } & = & - 2 \frac{J}{r} \sin^2{\theta} \\  
g_{rr} & = & 1 + 2 U + 4 U^2 - 2 \frac{Q}{r^3} P_2 
+ 2 \frac{J^2}{r^4} {(8 P_2-1)} - 10 \frac{M Q}{r^4} P_2 
+ \frac{1}{12} \frac{Q^2}{r^6} {(8 P_2^2 - 16 P_2 + 77)} 
\nonumber \\
g_{\theta \theta } & = & r^2 \left(1 -  2 \frac{Q}{r^3} P_2
+ \frac{J^2}{r^4} P_2 - 5 \frac{M Q}{r^4} P_2 
+ \frac{1}{36} \frac{Q^2}{r^6} (44 P_2^2 + 8 P_2 - 43) \right) \nonumber \\
g_{\phi \phi} & = & r^2 \sin^2{\theta} \left(1 - 2 \frac{Q}{r^3} P_2
+ \frac{J^2}{r^4} P_2 - 5 \frac{M Q}{r^4} P_2 
+ \frac{1}{36} \frac{Q^2}{r^6} (44 P_2^2 + 8 P_2 - 43) \right) , \nonumber
\end{eqnarray}

\noindent
where $ U = {M}/{r} $ and $ P_2 = ({3\cos^2{\theta} - 1})/{2} $. This new expanded HT form with second order quadrupole monent is a more convenient way to calculate the quantities we are going to obtain, because it is in Schwarzschild spherical coordinates. 


\section{The Geodesic Equation}
\label{geodesics}

The space-time interval between two events is defined as,

\begin{equation}
\label{ds}
ds^2 = g_{\alpha \beta} dx^{\alpha} dx^{\beta} .
\end{equation}

We can equate the interval with a proper time $ d \tau $ and so write down 
the following equation,

\begin{equation}
\label{mu}
\mu = g_{\alpha \beta} \frac{d x^{\alpha}}{d \tau} \frac{d x^{\beta}}{d \tau} ,
\end{equation}

\noindent
where $ \mu $ is a parameter to be defined. For massive particles moving across 
spacetime its trajectories are described by time-like intervals ($ ds^2 < 0 $), 
so we set $ \mu = +1 $, while light trajectories are described by light-like 
intervals ($ ds^2 = 0 $) and so we set $ \mu = 0 $.  
The former case is suitable for describing planetary motion, as its the case 
for planetary perihelion, while light deflection and time delay, which are 
light related, are described by the latter. The geodesic equations help to 
calculate the path with the shortest proper time between two points, 

\begin{equation}
\label{motion}
\frac{d}{d \tau} \left(g_{\alpha \beta} \frac{d x^{\beta}}{d \tau}\right) 
-\frac{1}{2} \partial_{\alpha} g_{\mu \nu} \frac{d x^{\mu}}{d \tau} 
\frac{d x^{\mu}}{d \tau} = 0 .
\end{equation}

The geodesic equation is related to conserved quantities, as in our case when 
we set $ \alpha = t $,

\begin{equation}
\label{const}
\frac{d}{d \tau}\left(g_{t t} \frac{d x^{t}}{d \tau}
+ g_{t \phi} \frac{d x^{\phi}}{d \tau}\right) = 0 .
\end{equation}

We can set the conserved quantity related with the energy $ E $, 

\begin{equation}
\label{econst}
g_{t t} \frac{d x^{t}}{d \tau} + g_{t \phi} \frac{d x^{\phi}}{d \tau} = - E .
\end{equation}

When we set $ \alpha = \phi $ we obtain a conserved quantity related to the 
density of angular momentum along the $ z $-axis, $ L_z $,

\begin{equation}
\label{lconst}
g_{\phi t} \frac{d x^{t}}{d \tau} + g_{\phi \phi} \frac{d x^{\phi}}{d \tau} = L_z .
\end{equation}

These relations can be reversed to obtain:

\begin{eqnarray}
\label{tphi1}
\frac{d t}{d \tau} & = & - \frac{1}{\rho^2} 
[- E g_{\phi \phi} - g_{t \phi} L_z] , \\
\label{tphi2}
\frac{d\phi}{d\tau} & = & - \frac{1}{\rho^2} [g_{tt} L_z + E g_{t \phi}] ,
\end{eqnarray}

\noindent
where $ \rho^2 = g_{t \phi}^2 - g_{\phi \phi} g_{t t} $. 
Equations (\ref{tphi1}) and (\ref{tphi2}) can be combined to,

\begin{equation}
\label{etphi}
\frac{d \phi}{d t} = \frac{d \phi}{d \tau} \frac{d \tau}{d t}
= \frac{g_{tt} L_z + E g_{t\phi}}{- E g_{\phi \phi} - g_{t \phi} L_z} .
\end{equation}


\section{Light Deflection}
\label{deflection}

\begin{figure}
\centering
\includegraphics[width=6cm]{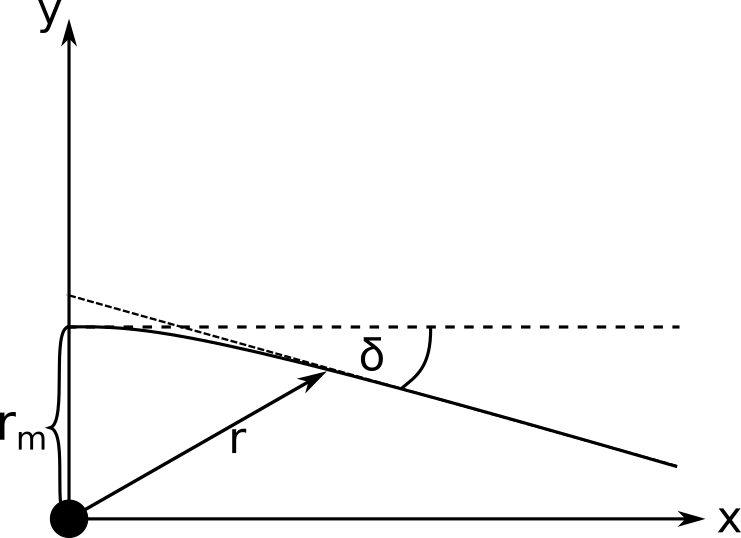}
\caption{Light deflection.}\label{Deflection}
\end{figure}

The effect is represented in Figure \ref{Deflection}. We set $ \mu = 0 $ in 
(\ref{mu}) and rearranging provides an equation for $ {d r}/{d t} $. 
We can use the substitution $ u = 1/r $  to obtain up to order 
$ O(M^2, \, Q^2, \, J^2) $:

\begin{eqnarray}
\label{ecdif}
\frac{d^2 u}{d \phi^2}& = & - 2 J \frac{E^3}{L_z^3} 
+ \left(12 J^2 \frac{E^4}{L_z^4}
- 8 J M \frac{E^3}{L_z^3} - 1 \right) u \nonumber\\
& + & \left(3 Q \frac{E^2}{L_z^2} + 3 M \right) u^2 \\
& + & \left(- 24 J Q \frac{E^3}{L_z^3} + 34 J^2 \frac{E^2}{L_z^2}
+ 10 M Q \frac{E^2}{L_z^2} \right) u^3 \nonumber\\
& + & \left(- \frac{81 J^2}{2} + \frac{3 M Q}{2} 
- \frac{93}{4} Q^2 \frac{E^2}{L_z^2} \right) u^5 + 33 Q^2 u^7 \nonumber
\end{eqnarray}

This equation can only be solved by perturbation theory. For this purpose, 
we propose a solution of the form  

\begin{eqnarray}
\label{ecu}
u & = & u_0 \cos{\phi} + c_0 u_m \nonumber \\
& + & J \left(u_0^3 u_{J1} + u_0^2 u_{J2} u_m + u_0 u_{J3} u_m^2 
+ u_{J4} u_m^3 \right) \nonumber \\
& + & M \left(u_0^2 u_{M1} + u_0 u_m u_{M2} + u_m^2 u_{M3} \right) \nonumber \\
& + & Q \left(u_0^4 u_{Q1} + u_0^3 u_m u_{Q2} + u_0^2 u_m^2 u_{Q3} + u_0 u_m^3 u_{Q4}
+ u_m^4 u_{Q5} \right) \nonumber \\
& + & J^2 \left(u_0^5 u_{JJ1} + u_0^4 u_{JJ2} u_m + u_0^3 u_{JJ3} u_m^2 \right. 
\nonumber \\
& + & \left. u_0^2 u_{JJ4} u_m^3 + u_0 u_{JJ5} u_m^4 + u_{JJ6} u_m^5 \right) 
\nonumber \\
& + & J M \left(u_0^4 u_{MJ1} + u_0^3 u_m u_{MJ2} + u_0^2 u_m^2 u_{MJ3} \right. 
\nonumber \\
& + & \left. u_0 u_m^3 u_{MJ4} + u_m^4 u_{MJ5} \right) \nonumber \\
& + & M^2 \left(u_0^3 u_{MM1} + u_0^2 u_m u_{MM2} + u_0 u_m^2 u_{MM3} + u_m^3 u_{MM4} 
\right) \nonumber \\
& + & M Q \left(u_0^5 u_{MQ1} + u_0^4 u_m u_{MQ2} + u_0^3 u_m^2 u_{MQ3} 
+ u_0^2 u_m^3 u_{MQ4} \right. \nonumber \\
& + & \left. u_0 u_m^4 u_{MQ5} + u_m^5 u_{MQ6} \right) \nonumber \\
& + & Q^2 \left(u_0^7 u_{QQ1} + u_0^6 u_m u_{QQ2} + u_0^5 u_m^2 u_{QQ3} 
+ u_0^4 u_m^3 u_{QQ4} \right. \nonumber \\
& + & \left. u_0^3 u_m^4 u_{QQ5} + u_0^2 u_m^5 u_{QQ6} + u_0 u_m^6 u_{QQ7} 
+ u_m^7 u_{QQ8} \right) \nonumber \\
& + & Q J \left(u_0^6 u_{QJ1}+u_0^5 u_m u_{QJ2}+u_0^4 u_m^2 u_{QJ3}
+ u_0^3 u_m^3 u_{QJ4} \right. \nonumber \\
& + & \left. u_0^2 u_m^4 u_{QJ5} + u_0 u_m^5 u_{QJ6} + u_m^6 u_{QJ7} \right) .
\end{eqnarray}

This method brings up a number of equations of the form:

\begin{equation}
\frac{d^2 u_{04}}{d \phi^2} = - u_{04} + 4 \frac{E^3}{L_z^3} \cos{\phi} ,
\end{equation}

\noindent
or,

\begin{equation}
\frac{d^2 u_{11}}{d \phi^2} = - u_{11} + 3 \cos^2{\phi} ,
\end{equation}

\noindent
and so on. For this part, we stuck to the general solutions to the 
differential equation as in \cite{BW},

$$ \frac{d^2 y}{d x^2} + y = \cos{(nx)} $$ 

\noindent
to be 

$$ y = - \frac{1}{n^2 - 1} \cos{(nx)} $$ 

\noindent
for $ n \neq 1 $ and 

$$ y = \frac{\phi}{2} \sin{\phi} $$ 

\noindent
for $ n = 1 $. The approximate solution is:

\begin{eqnarray}
\label{solecdif}
u & = & u_0 \cos{\phi} - 2 J u_m^3 
+ \frac{1}{2} M u_0^2 (3 - \cos{2 \phi}) \nonumber \\
& + &\frac{1}{2} Q u_0^2 u_m^2 (3 - \cos{2 \phi}) \nonumber \\
& - & \frac{81}{32} J^2 u_0^5 \left(5 \phi \sin{\phi} 
+ \frac{5}{8} \cos{3 \phi} + \frac{1}{24} \cos{5 \phi} \right) \nonumber \\
& + & J^2 u_0^3 u_m^2 \left(\frac{51}{4} \phi  \sin{\phi}
+ \frac{17}{16} \cos{3 \phi} \right) \nonumber \\
& + & 6 J^2 u_0 u_m^4 \phi \sin{\phi} \nonumber \\
& + & 2 J M u_0 u_m^3 \phi \sin{\phi} \nonumber \\
& + &J Q u_0^3 u_m^3 \left(\frac{3}{4} \cos{3 \phi} 
- 9 \phi \sin{\phi} \right)\nonumber \\
& - & 6 J Q u_0 u_m^5 \phi \sin{\phi} \nonumber \\
& + & M^2 u_0^3 \left(\frac{15}{4} \phi \sin{\phi}
-\frac{3}{16} \cos{3 \phi} \right)\nonumber \\
& + & M Q u_0^5 \left(\frac{15}{2} \phi \sin{\phi}
+ \frac{15}{16} \cos{3 \phi}+\frac{1}{16} \cos{5 \phi} \right) \nonumber \\
& + & M Q u_0^3 u_m^2 \left(\frac{45}{4} \phi  \sin{\phi}
- \frac{1}{16} \cos{3 \phi} \right) \nonumber \\
& + & \frac{33}{64} Q^2 u_0^7 \left(\frac{35}{2} \phi \sin{\phi}
+ \frac{21}{8} \cos{3 \phi} \right. \nonumber \\
& + & \left. \frac{7}{24} \cos{5 \phi}
+ \frac{1}{48} \cos{7 \phi}\right) \nonumber \\
& - &\frac{93}{64} Q^2 u_0^5 u_m^2 \left(5 \phi \sin {\phi} 
+ \frac{5}{8} \cos{3 \phi} +\frac{1}{24} \cos{5 \phi}\right) \nonumber \\
& + &Q^2 u_0^3 u_m^4 \left(\frac{15}{4} \phi \sin{\phi} 
-\frac{3}{16} \cos{3 \phi} \right) 
\end{eqnarray}

The closest approach $ u_m $ occurs when $ \phi = 0 $, so:
 
\begin{eqnarray}
\label{closest}
u_m & = & u_0 - 2 J u_m^3 + M u_0^2 + Q u_0^2 u_m^2 \nonumber \\
& - & J^2 u_0^3 \left(\frac{27 u_0^2}{16} + \frac{17}{16} u_m^2 \right) 
\nonumber \\
& + & \frac{3}{4} J Q u_0^3 u_m^3-\frac{3 M^2 u_0^3}{16} \nonumber \\
& + & M Q u_0^3 \left(u_0^2 - \frac{1}{16} u_m^2 \right) \nonumber \\
& + & Q^2 u_0^3 \left(\frac{1551 u_0^4}{1024}-\frac{31}{32} u_0^2 u_m^2 
- \frac{3}{16} u_m^4 \right)
\end{eqnarray}

The deflection angle $ \Delta \phi = 2\delta $ can be found using the condition 
$ u(\pi/2 + \delta) = 0 $, that is:

\begin{eqnarray}
\label{lightdef}
\Delta \phi & = & 4 M u_m - 4 J u_m^2 + 4 Q u_m^3 \nonumber \\ 
& + & \left(8 + \frac{195}{32} \pi \right) J^2 u_m^4
+ 2 \left(2 + \pi \right) J M u_m^3\nonumber \\ 
& + & \left(4 - 15 \pi \right) J Q u_m^5 
- \left(4 - \frac{15}{4} \pi \right) M^2 u_m^2 \nonumber \\
& - & \left(8 - \frac{75}{4} \pi \right) M Q u_m^4 \nonumber \\
& + & \left(\frac{705}{128} \pi - 4 \right) Q^2 u_m^6 .
\end{eqnarray}

This result agrees with the result expected from the Schwarzschild metric, 

$$ \Delta \phi \approx 4 M u_m 
- \left(4 - \frac{15}{4} \pi \right) M^2 u_m^2 , $$

\noindent 
up to second order in mass \cite{BW}. The evaluation of some these terms for 
a ray of light grazing the solar limb is presented in Table \ref{table:3}

\begin{table}[ht!]
\centering
\begin{tabular}{|c c c c c|} 
\hline
Stellar Object & First Order: Mass ($\mu$as) &Rotation ($\mu$as) & 
Second Order: Mass ($\mu$as) & Quadrupole ($\mu$as)  \\ [0.5ex] 
\hline\hline
Sun & $1.75175\times 10^6$ &$6.991859 \times 10^{-1}$ & 7.224014 &  9.627369 \\ 
Mercury & $8.292245 \times 10^1$ & $3.287143 \times 10^{-7}$
& $1.621187\times 10^{-8}$& -  \\
Venus & $4.929369\times 10^2$ & $1.011539 \times 10^{-6}$ 
& $5.728902 \times 10^{-7}$&  -\\
Earth & $5.736892 \times 10^2$&$ 2.960172 \times 10^{-4}$ 
& $7.759650 \times 10^{-7}$& $6.210989 \times 10^{-1}$ \\
Mars & $1.158410 \times 10^2$ & $3.490819 \times 10^{-5}$
& $3.163833 \times 10^{-8}$ & $2.275118 \times 10^{-1}$ \\
Jupiter & $1.641520 \times 10^4$ & $1.705688 \times 10^{-1}$
& $ 6.353035 \times 10^{-4}$& $2.421242 \times 10^{2}$ \\
Saturn & $5.802427\times 10^3$  & $4.320800 \times 10^{-2}$ 
& $7.937946 \times 10^{-5}$& $9.544993 \times 10^{-1}$ \\
Uranus & $2.172504 \times 10^3$ & $3.336672 \times 10^{-3}$
& $1.112782 \times 10^{-5}$ & $2.607005 \times 10^{-1}$ \\
Neptune & $2.508570 \times 10^3$ & $8.357399 \times 10^{-3}$
& $1.483684 \times 10^{-5}$& $1.003428 \times 10^{-1}$ \\[1ex] 
\hline
\end{tabular}
\caption{Order of magnitude of some of the contributions to the deviation angle 
of a light ray grazing the solar limb as predicted by our model.}
\label{table:3}
\end{table}


\section{Precession of the Perihelion} 
\label{perihelion}

The effect is represented in Figure \ref{Peri}. First, we use the geodesic 
equation (\ref{motion}) to find the conserved quantities, and the equations 
(\ref{tphi1}) and (\ref{tphi2}). Using these new identities, it is possible to 
calculate $ {d r}/{d \tau} $ setting $ \mu = 1 $ in (\ref{mu}) and imposing a 
planar orbit $ (\theta = \pi/2) $. After this, the well known variable change 
$ u = 1/r $ is used, so it is possible to find $ u = u(\phi) $ by means of:

\begin{equation}
\frac{d u}{d \phi} = \frac{d u}{d \tau} \frac{d \tau}{d \phi} . 
\end{equation}

After taking the second derivative with respect to $ \phi $, we found up 
to order $ O(M^2, \, Q^2, \, J^2)$, the result is:

\begin{figure}
\centering
\includegraphics[width=5cm]{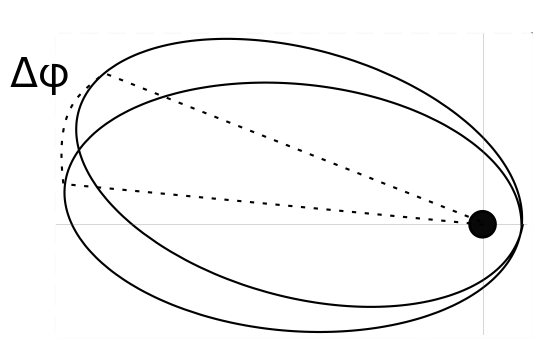}
\caption{Perihelion of a planet.} \label{Peri}
\end{figure}

\begin{eqnarray}
\frac{d^2 u}{d \phi^2} & = & 2 J \frac{E}{L_z^3} - 2 J \frac{E^3}{L_z^3} 
+ \frac{M}{L_z^2} \nonumber \\
& + & u \left(12 J^2 \frac{E^4}{L_z^4} - 8 J M \frac{E^3}{L_z^3}
- 12 J^2 \frac{E^2}{L_z^4}-1\right) \nonumber \\
& + & u^2 \left(3 M + 3 Q^2 \frac{E^2}{L_z^2} - \frac{3}{2} \frac{Q}{L_z} \right)
\nonumber \\
& + & u^3 \left(- 24 J Q \frac{E^3}{L_z^3} + 34 J^2 \frac{E^2}{L_z^2}
+ 10 M Q \frac{E^2}{L_z^2} \right. \nonumber \\
& + & \left. 16 J Q \frac{E}{L_z^2}-34 \frac{J^2}{L_z^2}\right)
\nonumber \\
& + & u^5 \left(- \frac{93}{4} Q^2 \frac{E^2}{L_z^2}
- \frac{81}{2} J^2 + \frac{111}{4}\frac{Q^2}{L_z^2}
+ \frac{3}{2} M Q \right) \nonumber \\
& + & 33 Q^2 u^7
\end{eqnarray}

We can consider a perturbation $ u = u_c + u_c w(\phi) $, where $ w $ is 
the wobble function we want to find. As such, given that $ w << 1$, 
it satisfies the harmonic equation:

\begin{eqnarray}
\frac{d^2 w}{d \phi^2} + w & = &
\left(6 \frac{M}{r_c} + 3 \frac{Q}{L_z^2r_c} \left(2E^2 - 1\right)
+ 3 J^2 \left(\frac{4E^4-4E^2}{L_z^4} + \frac{34E^2-34}{L_z^2 r_c^2} 
- \frac{135}{2r_c^4} \right) \right. \nonumber \\
& - & \left. 8 E^3 \frac{J M}{L_z^3} + 24 E\frac{J Q}{L_z^3r_c^2} \left(2 
- 3 E^2 \right)
\right. \nonumber \\ 
& + & \left.\frac{15}{2} \frac{M Q}{r_c^2}\left(\frac{4E^2}{L_z^2}  
+ \frac{1}{r_c^2} \right)
+ \frac{3}{4} \frac{Q^2}{r_c^4} \left(\frac{185 - 155 E^2}{L_z^2} 
+ \frac{308}{r_c^2} \right) \right) w .
\end{eqnarray}

It provides an angular frequency $ \omega $ value for which 
$ w = A \cos{(\omega \phi + \phi_0)} $. The orbit perihelion $ \Delta \phi $ 
occurs when $ w(\phi) $ is a minimum, i.e. when the argument of the cosine 
function is $ \pi + 2\pi n $.  $ \Delta \phi $ can be found using the condition 
$ \omega \Delta \phi = 2 \pi $.  Although other methods can be used 
\cite{D'inverno}, by using the common substitution 

$$ \hat{E} = \frac{E^2-1}{2} , $$

\noindent
along the Schwarzschild circular orbit approximation 

$$ \hat{E} \approx -\frac{M}{r_c} + \frac{L_z^2}{2r_c^2} 
- \frac{M L_z^2}{r_c^3} $$ 

\noindent
this implies:

\begin{eqnarray}
\Delta \phi & = & 6 \pi \frac{M}{r_c}
+ 3 \pi \frac{Q}{r_c} \left(\frac{1}{L_z^2} + \frac{2}{r_c^2} \right)
\nonumber \\   
& - & 3 \pi \frac{J^2}{r_c^2} \left(\frac{4}{L_z^2}
+ \frac{59}{2 r_c^2} \right) \nonumber \\  
& - & 8 \pi \frac{J M}{L_z r_c} \sqrt{L_z^2+r_c^2} \left(\frac{1}{L_z^2} 
+ \frac{1}{r_c^2}\right)
+ 24 \pi \frac{J Q}{L_z r_c^3} \sqrt{L_z^2+r_c^2} \left(\frac{1}{L_z^2} 
+ \frac{3}{r_c^2} \right)
\nonumber \\ 
& + & 27\pi \frac{M^2}{r_c^2}  
+ 3 \pi \frac{M Q}{2 r_c^2} \left(\frac{30}{L_z^2} + \frac{53}{ r_c^2} \right)
\nonumber \\  
& + & \frac{9}{4} \pi \frac{Q^2}{r_c^2} \left(\frac{3}{L_z^4} 
+ \frac{22}{L_z^2 r_c^2} + \frac{63}{r_c^4} \right)
\end{eqnarray} 

This result agrees with the result expected from the Schwarzschild metric, 
$ \Delta \phi \approx 6 \pi {M}/{r_c} $, up to first order in mass. 
For the perihelion precession of Mercury some of the contributions can be 
computed as is shown in Table \ref{table:Precession}. The gravitational 
periastron precession in the orbit of the star S2 are also included, and they 
agree whit the reported value in literature of 12 arcmin per orbit 
($\approx$ 75 arcsec per century) near the pericentre \cite{Gravity}.

\begin{table}[ht!]
\centering
\begin{tabular}{|c c c c|} 
\hline
Body & First Order: Mass (as/cent) &Second Order: Mass (as/cent) 
& Quadrupole (as/cent)   \\ [0.5ex] 
\hline\hline
 S2 & 73.8075 & 0.00101207 & -\\
 Mercury &41.162 & 4.72301$\times$10$^{-6}$ &4.72301$\times $10$^{-6}$\\[1ex]
\hline
\end{tabular}
\caption{Order of magnitude of the contributions to the gravitational 
periastron and  perihelion precessions in the orbits of the star S2 and 
Mercury, respectively.}
\label{table:Precession}
\end{table}


\section{Time Delay}
\label{delay}

\begin{figure}[h]
\centering
\includegraphics[width=4.5cm]{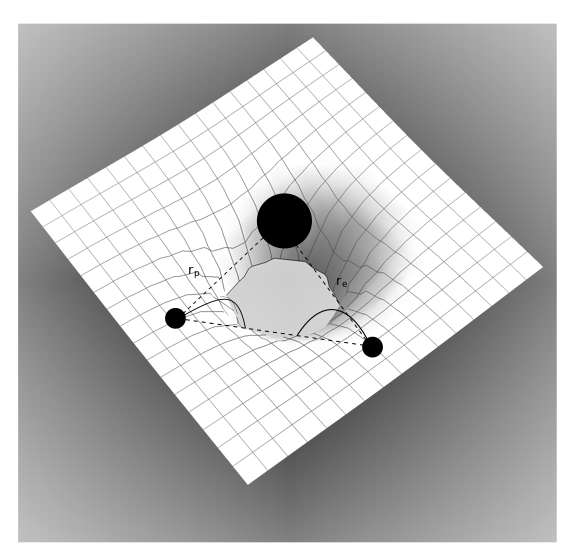}
\caption{Time delay of light signals.} \label{TD}
\end{figure}

The effect is represented in Figure \ref{TD}, as the path of rays of light are 
turned away from their classical trajectories. The curvature induced in the 
spacetime surrounding a massive body increases the travel time of light rays 
relative to what would be the case in flat space. Let $ b $ be the maximum 
approach distance of a ray of light traveling near a massive body. If the beam 
traveled in a straight line, then $ r \cos{\phi} = b $. This means that

$$ d \phi = \frac{b dr}{\sqrt{r^2 - b^2}} . $$

By using $d \theta = 0 $, it is possible to extract $ d t $ from 
$ g_{\mu \nu} dx^{\mu}dx^{\nu} = 0 $, so we obtain:

\begin{eqnarray}
d t & = & \frac{d r}{\sqrt{r^2 - b^2}} \left[2 M + r - 2 \frac{J b}{r^2} 
- \frac{M b^2}{r^2} + \frac{Q}{r^2} \right. \nonumber \\
& - & \left. 5 \frac{J^2}{r^3} + \frac{27}{4} \frac{J^2 b^2}{r^5} 
- 4 \frac{J M b}{r^3} - 2 \frac{J Q b}{r^5} \right. \nonumber \\
& + & \left. 4 \frac{M^2}{r} - 2 \frac{M^2 b^2}{r^3} 
- \frac{1}{2} \frac{M^2 b^4}{r^5}  + 5 \frac{M Q}{r^3} 
- \frac{5}{4} \frac{M Q b^2}{r^5} \right. \nonumber \\
& + & \left. \frac{31}{8} \frac{Q^2}{r^5} 
- \frac{33}{8} \frac{Q^2 b^2}{r^7} \right] .
\end{eqnarray}%
  
Performing an integration to go from a planet at position $ r_e $, to another 
planet at $ r_p $, to find the time delay:

\begin{eqnarray}
\Delta t & = & d_e + d_p \nonumber \\ 
& + & 2 J \left(\frac{b}{r_e d_e} + \frac{b}{r_p d_p} 
- \frac{r_e}{b d_e} - \frac{r_p}{b d_p} \right) \nonumber \\
& + & 2 M \log{\left[\frac{(r_e+d_e)(r_p+d_p)}{b^2}\right]} \nonumber \\ 
& + & M \left(\frac{b^2}{r_e d_e} + \frac{b^2}{r_p d_p} 
- \frac{r_e}{d_e} - \frac{r_p}{d_p} \right) \nonumber \\
& + & Q \left(\frac{d_e}{b^2 r_e} + \frac{d_p}{b^2 r_p} \right) \nonumber \\
& - & \frac{27}{16} J^2 \left(\frac{b^2}{r_e^4 d_e} 
- \frac{b^2}{r_p^4 d_p} \right)
+ \frac{53}{32} J^2 \left(\frac{1}{r_e^2 d_e} 
+ \frac{1}{r_p^2 d_p} \right) \nonumber \\
& + & \frac{1}{32} J^2 \left(\frac{\pi }{b^3} 
- \frac{\theta_e}{b^3} - \frac{\theta_p}{b^3} 
+ \frac{1}{b^2 d_e} + \frac{1}{b^2 d_p} \right) \nonumber\\
& + & 2 J M \left(\frac{b}{r_e^2 d_e} + \frac{b}{r_p^2 d_p} 
- \frac{1}{b d_e} - \frac{1}{b d_p} \right) \nonumber \\
& + & 2 J M \left(\frac{\theta_e}{b^2} + \frac{\theta_p}{b^2} 
- \frac{\pi}{b^2} \right) \nonumber \\
& + & \frac{1}{2} J Q \left(\frac{b}{r_e^4 d_e} + \frac{b}{r_p^4 d_p} \right) 
+ \frac{3}{4} J Q \left(\frac{\theta_e}{b^4} + \frac{\theta_p}{b^4} \right) 
\nonumber \\ 
& - & \frac{3}{4} J Q \left(\frac{1}{b^3 d_e} + \frac{1}{b^3 d_p} 
+ \frac{\pi}{b^4} \right)
+ \frac{1}{4} J Q \left(\frac{1}{b r_e^2 d_e} + \frac{1}{b r_p^2 d_p} \right) 
\nonumber \\
&+& \frac{1}{8} M^2 \left(\frac{b^4}{r_e^4 d_e} + \frac{b^4}{r_p^4 d_p} \right) 
+ \frac{9}{16} M^2 \left(\frac{b^2}{r_e^2 d_e} + \frac{b^2}{r_p^2 d_p} \right) 
\nonumber \\
& + & \frac{37}{16} M^2 \left(\frac{\pi}{b} 
- \frac{\theta_e}{b} - \frac{\theta_p}{b} \right)
- \frac{11}{16} M^2 \left(\frac{1}{d_e} + \frac{1}{d_p} \right) \nonumber \\
& + & \frac{5}{16} M Q \left(\frac{b^2}{r_e^4 d_e} + \frac{b^2}{r_p^4 d_p} 
\right) \nonumber \\
& + & \frac{65}{32} M Q \left(\frac{\pi}{b^3} 
- \frac{\theta_e}{b^3} - \frac{\theta_p}{b^3} 
+ \frac{1}{b^2 d_e} + \frac{1}{b^2 d_p} \right) \nonumber \\
& - & \frac{75}{32} M Q \left(\frac{1}{r_e^2 d_e} + \frac{1}{r_p^2 d_p} \right) 
\nonumber \\ 
& + & \frac{11}{16} Q^2 \left(\frac{b^2}{r_e^6 d_e} +\frac{b^2}{r_p^6 d_p} 
\right) \nonumber \\
& + & \frac{21}{128} Q^2 \left(\frac{\pi}{b^5} 
- \frac{\theta_e}{b^5} - \frac{\theta_p}{b^5} + \frac{1}{b^4 d_e} 
+ \frac{1}{b^4 d_p} \right) \nonumber \\
& - & \frac{7}{128} Q^2 \left(\frac{1}{b^2 r_e^2 d_e} 
+ \frac{1}{b^2 r_p^2 d_p} \right) \nonumber \\
& - & \frac{51}{64} Q^2 \left(\frac{1}{r_e^4 d_e} + \frac{1}{r_p^4 d_p} \right) 
\end{eqnarray}

\noindent
where $ d_e = \displaystyle{\sqrt{r_e^2 - b^2}} $, $ d_p = \sqrt{r_p^2 - b^2} $, 
$ \theta_e = \sin^{-1}({b}/{r_e}) $, and $ \theta_p = \sin^{-1}({b}/{r_p}) $.

This result agrees with the result expected from the Schwarzschild metric, 
up to first order in mass \cite{McMahonDemystified}. Some of the contributions 
of the gravitational delay of light grazing the solar limb and the planets as 
predicted by our model are presented in Table \ref{table:4}.

\begin{table}[ht!]
\centering
\begin{tabular}{|c c c c c|} 
\hline
Stellar Object & Second Order: PN (ns) &Rotation (ns) & 
Fourth Order: PPN (ns) & Quadrupole: PN (ns)  \\ [0.5ex] 
\hline\hline
Sun & $1.096102\times 10^6$ &$7.869329 \times 10^{-3}$ 
& $3.033948 \times 10^{-1}$ &  $5.417914 \times 10^{-2}$ \\ 
Mercury & $3.508723 \times 10^{-2}$ & $1.296546\times 10^{-11}$
& $2.388131\times 10^{-12}$& -  \\
Venus & $4.352088\times 10^{-1}$ & $9.896817 \times 10^{-11}$ 
& $2.093294 \times 10^{-10}$&  -\\
Mars & $6.510764 \times 10^{-2}$ & $1.916006\times 10^{-9}$
& $ 6.485408 \times 10^{-12}$& $6.243720 \times 10^{-6}$ \\
Jupiter & $1.746187\times 10^2$  & $1.954326 \times 10^{-4}$ 
& $2.718487 \times 10^{-6}$& $1.387094 \times 10^{-1}$ \\
Saturn & $5.722455 \times 10^1$ & $4.192500 \times 10^{-5}$
& $2.876544 \times 10^{-7}$ & $4.630785 \times 10^{-2}$ \\
Uranus & $1.016421\times 10^1$ & $1.322018 \times 10^{-6}$
& $1.646610 \times 10^{-8}$& $5.164588 \times 10^{-3}$ \\
Neptune & $1.248393\times 10^1$ & $3.392365\times 10^{-6}$
& $2.249210 \times 10^{-8}$& $2.036515 \times 10^{-3}$ \\[1ex] 
\hline
\end{tabular}
\caption{Order of magnitude of the contributions PN, PPN, $ {\rm PN}_{\rm Q} $ 
and $ {\rm PN}_{\rm R} $ to the time delay of a light ray grazing the solar 
limb as predicted by our model.}
\label{table:4}
\end{table}


\section{Gravitational Redshift}
\label{redshift}

\begin{figure}[h!]
\centering
\includegraphics[width=6cm]{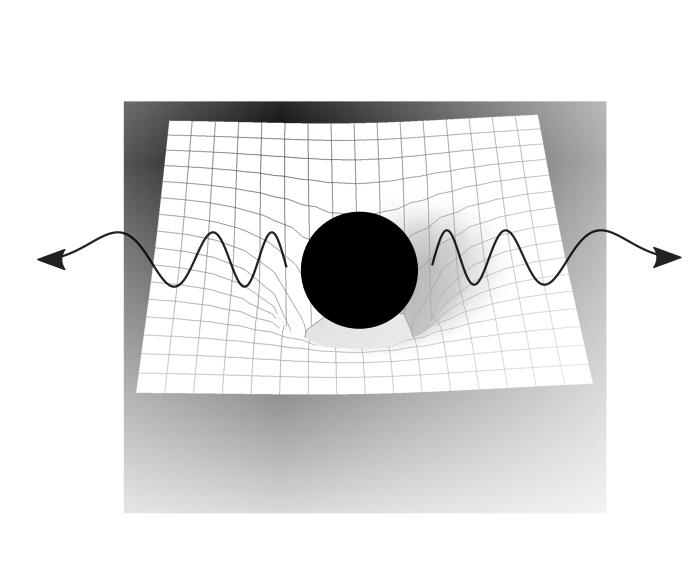}
\caption{Gravitational redshift.}\label{RShift}
\end{figure}

The effect is represented in Figure \ref{RShift}. It is possible to calculate 
a redshift factor by comparing the proper time for observers located at two 
different values of $r$, assuming a planar orbit, $ \theta = \pi/2 $. 

\begin{eqnarray}
\frac{\lambda_r}{\lambda_e} & = & \sqrt{\frac{g_{tt}(r_r)}{g_{tt}(r_e)}} \approx 
1 + M \left(\frac{1}{r_e} - \frac{1}{r_r} \right)
+ \frac{Q}{2} \left(\frac{1}{r_e^3} - \frac{1}{r_r^3} \right) \nonumber \\
& + & \frac{3}{2} M^2 \left(\frac{1}{r_e^2} - \frac{2}{r_er_r} 
-\frac{1}{r_r^2} \right) \nonumber \\
& + & M Q \left(\frac{2}{ r_e^4}-\frac{1}{2 r_e^3 r_r}
- \frac{1}{2  r_e r_r^3} - \frac{1}{r_r^4} \right) \nonumber \\
& + & Q^2 \left(\frac{1}{8 r_e^6} - \frac{1}{4 r_e^3 r_r^3} 
+ \frac{1}{8 r_r^6} \right)
\end{eqnarray}

This result agrees with the result expected from the Schwarzschild metric, 
up to first order in mass \cite{MooreTA}. The gravitational redshift in 
the orbit of the star S2 agrees with the reported value in literature of 
103 km s$^{-1}$/c near the pericentre \cite{Gravity, Zucker}, as it is shown 
in Table \ref{table:5}.

\begin{table}[ht!]
\centering
\begin{tabular}{|c c|} 
\hline
First Order: Mass (km s$^{-1}$/c) & 
Second Order: Mass (km s$^{-1}$/c)    \\ [0.5ex] 
\hline\hline 
103.24 & 0.0532923 \\ [1ex] 
\hline
\end{tabular}
\caption{Order of magnitude of the contributions to the gravitational redshift 
in the orbit of the star S2.}
\label{table:5}
\end{table}


\section{Conclusions}
\label{conclusions}

We reviewed the calculations of the classical experiments in GR with an 
approximative metric and taking in account all second order terms of mass,
angular momentum and mass quadrupole. If we neglect these terms our results 
agree with the ones in the literature. By using our results, it could now be
possible to estimate the value of second order terms of mass, quadrupole and
angular momentum and determine how well they adapt to the predicted phenomena
in the classical tests.

In PPN theory these results were obtained, but in this theory the quadrupole 
moment is introduced in the expansion of the mass potential. Here, this effect 
is introduced by the metric in a straightforward way. Our calculations 
were done in a simple manner using Mathematica. Moreover, we developed a 
Mathematica notebook, which is available upon request. The notebook is divided 
in sections, each one correspoding to a classical test. These calculations in 
the PPN method are rather complicated, but it would be interesting to expand 
them using the PPN methods.

As future work, it is planned to include the spin octupole and 
the mass hexadecapole, because now, these relativistic multipoles are 
currently considered in neutron stars calculations. For instance, to determine
the innermost stable circular orbit or the precession frequencies,
these relativistic multipole moment play an important role \cite{Ryan,Shibata}. 
Moreover, it would be interesting to investigate the effect of the quadrupole
moment in the gravitational lens effect. To do it, one has to employ
the PPN formalism. The results of this research can also serve as a basis for 
predicting the effects of rotation when better MBH spin measurements have been 
made.



%

\end{document}